\documentstyle[12pt]{article}

\begin{document}

\hoffset = -1truecm
\voffset = -2truecm
\baselineskip = 10 mm

\title{\bf Parton recombination effect in polarized parton distributions }

\author{
{\bf Wei Zhu},
{\bf Zhenqi Shen} and {\bf Jianhong Ruan} \\
\normalsize Department of Physics, East China Normal University,
Shanghai 200062, P.R. China \\
}

\date{}

\newpage

\maketitle

\vskip 3truecm

\begin{abstract}

    Parton recombination corrections to the standard spin-dependent
Altarelli-Parisi evolution equation are considered in a nonlinear
evolution equation. The properties of this recombination equation
and its relation with the spin-averaged form are discussed.
\end{abstract}


$keywords$: Parton recombination; PQCD evolution equation; spin of
nucleon

\newpage
\begin{center}
\section{Introduction}
\end{center}

    The spin problem of the nucleon has been investigated
extensively since the discovery of the EMC spin effect, which
indicates that a small fraction of the nucleon spin is from the
quark helicity [1]. This surprising result leads to extensive
study on the gluon- and sea quark-polarization. The current
parameterizations suggest the non-vanished polarization in gluon.
It implies that the spin-correlation exists among polarized
partons. Determination of polarized parton distribution functions
is crucial for understanding the spin structure of the nucleon.
The knowledge about polarized parton distributions in the nucleon
is mainly extracted from global analysis of polarized experimental
data at different values of $Q^2$ based on the linear (spin
dependent) Altarelli-Parisi (AP) equation [2]. However, the
contributions of spin-correlation among initial polarized partons
are excluded in the linear evolution equation. Therefore, it is
necessary to modify the AP equation to include such correlation
effect.

    In the unpolarized case, the parton recombination as a leading
correlation effect of partons in QCD evolution equation has been
studied at leading logarithmic ($Q^2$) approximation ($LLA(Q^2)$)
and in a whole $x$ region by [3-4]. We call it the recombination
evolution equation. The parton recombination effect is generally
considered as a phenomenon at the small x region, where the rapid
increase of parton densities can be obviously suppressed, i.e.,
the saturation is predicted [5]. The behavior of polarized parton
distributions at small x is an important issue because the
experimental determination of the first moment of polarized gluon
distribution depending on it is rather strongly [1,6]. Moreover,
the parton recombination can also influence the polarized parton
distributions at a larger $x$ region. For example, the helicity of
a fast gluon may be changed after combining with a slower gluon,
although its momentum is almost unchanged. Therefore, in this
paper we generalize the recombination equation to the
spin-dependent case. This equation can work in a whole $x$ region.
Thus, the recombination equation can be used to study the spin
effect. In particular, we show that the polarized recombination
effect disappears at the small $x$ limit. We also find that
different from the linear evolution equation, the correlation form
of polarized partons in the spin-dependent recombination is
relevant to the structure of the spin-independent evolution
kernel. Therefore, this research is also useful for understanding
the shadowing dynamics in the unpolarized parton distributions.

    The paper is organized as follows. In Section 2 we construct the
spin-dependent recombination equation using the same way as refs.
[3,4]. In Section 3 the polarized recombination functions are
evolute. Then the discussion and concluding remarks are given in
Section 4.

\newpage
\begin{center}
\section{Recombination evolution equation}
\end{center}

    The elementary process is one-parton splitting to
two-partons in the AP equation. At very large number densities of
partons, for example in the small $x$ region, the wave functions
of partons can overlap. In this case the contributions of
two-partons-to-two-partons subprocess (i.e., the parton
recombination) should be considered in the QCD evolution
equations. A complete derivation for such equations need to sum
the contributions from real and interference Feynman diagrams and
corresponding virtual diagrams.  Gribov, Levin and Ryskin in [7]
use the AGK cutting rule [9] to count the contributions of
inteference diagrams. On the other hand, the real diagrams are
calculated at the double leading logarithmic approximation
($DLLA$) in a covariant way (the cut vertex technique) by Mueller
and Qiu in [8]. This nonlinear evolution is called the GLR-MQ
equation.

    Unfortunately, the application of the AGK cutting rule in the
GLR-MQ equation was questioned since it breaks the evolution
kernels [3]. To avoid this disadvantage, the simple relations
among the relevant high twist amplitudes are derived by using the
time ordered perturbation theory (TOPT) at the leading logarithmic
approximation ($LLA(Q^2)$) in [3]. Then the recombination
functions are calculated in a whole $x$ region based on the same
TOPT-framework in [4]. This is the recombination evolution
equation.

     The helicity amplitudes in the recombination
equation satisfy the same TOPT-relations as in the unpolarized
case. Thus, using the arguments in [3] we can directly write down
the spin-dependent evolution equations which are

$$\frac{dxG_\pm(x,Q^2)}{d\ln Q^2}$$
$$=\frac{\alpha_s^2K}{Q^2}
\int_{x_1+x_2\ge x}dx_1dx_2[Fig.1a+Fig.1b+Fig.1c]$$
$$-2\int_{(x_1+x_2)/2\ge x}dx_1dx_2[Fig.1a+Fig.1b+Fig.1c],\eqno(1a)$$

$$Fig.1a=f_{G_+G_+}(x_1,x_2, Q^2) R_{G_+G_+\rightarrow G_\pm }(x_1,x_2,x)$$
$$+f_{G_+G_-}(x_1,x_2, Q^2)R_{G_+G_-\rightarrow G_\pm }(x_1,x_2,x)$$
$$+f_{G_-G_+}(x_1,x_2, Q^2)R_{G_-G_+\rightarrow G_\pm }(x_1,x_2,x)$$
$$+f_{G_-G_-}(x_1,x_2, Q^2)R_{G_-G_-\rightarrow G_\pm }(x_1,x_2,x),
\eqno(1b)$$

$$Fig.1b=f_{q^i_+\overline{q}^i_+}(x_1,x_2, Q^2) R_{q^i_+\overline{q}^i_+\rightarrow G_\pm }(x_1,x_2,x)$$
$$+f_{q^i_+\overline{q}^i_-}(x_1,x_2, Q^2)R_{q^i_+\overline{q}^i_-\rightarrow G_\pm }(x_1,x_2,x)$$
$$+f_{q^i_-\overline{q}^i_+}(x_1,x_2, Q^2)R_{q^i_-\overline{q}^i_+\rightarrow G_\pm }(x_1,x_2,x)$$
$$+f_{q^i_-\overline{q}^i_-}(x_1,x_2, Q^2)R_{q^i_-\overline{q}^i_-\rightarrow G_\pm }(x_1,x_2,x),
\eqno(1c)$$

$$Fig.1c=f_{G_+q^i_+}(x_1,x_2, Q^2) R_{G_+q^i_+\rightarrow G_\pm }(x_1,x_2,x)$$
$$+f_{G_+q^i_-}(x_1,x_2, Q^2)R_{G_+q^i_-\rightarrow G_\pm }(x_1,x_2,x)$$
$$+f_{G_-q^i_+}(x_1,x_2, Q^2)R_{G_-q^i_+\rightarrow G_\pm }(x_1,x_2,x)$$
$$+f_{G_-q^i_-}(x_1,x_2, Q^2)R_{G_-q^i_-\rightarrow G_\pm }(x_1,x_2,x),
\eqno(1d)$$

$$\frac{dxq^i_\pm(x,Q^2)}{d\ln Q^2}$$
$$=\frac{\alpha_s^2K}{Q^2}
\int_{x_1+x_2\ge x}dx_1dx_2[Fig.1d+Fig.1(e-1)+Fig.1(e-2)+Fig.
1f]$$
$$-2\int_{(x_1+x_2)/2\ge x}dx_1dx_2[Fig.1d+Fig.1(e-1)+Fig.1(e-2)+Fig.
1f],\eqno(1e)$$

$$Fig.1d=f_{G_+G_+}(x_1,x_2, Q^2) R_{G_+G_+\rightarrow q^i_\pm }(x_1,x_2,x)$$
$$+f_{G_+G_-}(x_1,x_2, Q^2)R_{G_+G_-\rightarrow q^i_\pm }(x_1,x_2,x)$$
$$+f_{G_-G_+}(x_1,x_2, Q^2)R_{G_-G_+\rightarrow q^i_\pm }(x_1,x_2,x)$$
$$+f_{G_-G_-}(x_1,x_2, Q^2)R_{G_-G_-\rightarrow q^i_\pm }(x_1,x_2,x),
\eqno(1f)$$

$$Fig.1(e-1)=f_{q^j_+\overline{q}^j_+}(x_1,x_2, Q^2) R_{q^j_+\overline{q}^j_+\rightarrow q^i_\pm }(x_1,x_2,x)$$
$$+f_{q^j_+\overline{q}^j_-}(x_1,x_2, Q^2)R_{q^j_+\overline{q}^j_-\rightarrow q^i_\pm }(x_1,x_2,x)$$
$$+f_{q^j_-\overline{q}^j_+}(x_1,x_2, Q^2)R_{q^j_-\overline{q}^j_+\rightarrow q^i_\pm }(x_1,x_2,x)$$
$$+f_{q^j_-\overline{q}^j_-}(x_1,x_2, Q^2)R_{q^j_-\overline{q}^j_-\rightarrow q^i_\pm }(x_1,x_2,x),
\eqno(1g)$$

$$Fig.1(e-2)=f_{q^j_+q^k_+}(x_1,x_2, Q^2) R_{q^j_+q^k_+\rightarrow q^i_\pm }(x_1,x_2,x)$$
$$+f_{q^j_+q^k_-}(x_1,x_2, Q^2)R_{q^j_+q^k_-\rightarrow q^i_\pm }(x_1,x_2,x)$$
$$+f_{q^j_-q^k_+}(x_1,x_2, Q^2)R_{q^j_-q^k_+\rightarrow q^i_\pm }(x_1,x_2,x)$$
$$+f_{q^j_-q^k_-}(x_1,x_2, Q^2)R_{q^j_-q^k_-\rightarrow q^i_\pm }(x_1,x_2,x),
\eqno(1h)$$

$$Fig.1f=f_{q^i_+G_+}(x_1,x_2, Q^2) R_{q^i_+G_+\rightarrow q^i_\pm }(x_1,x_2,x)$$
$$+f_{q^i_+G_-}(x_1,x_2, Q^2)R_{q^i_+G_-\rightarrow q^i_\pm }(x_1,x_2,x)$$
$$+f_{q^i_-G_+}(x_1,x_2, Q^2)R_{q^i_-G_+\rightarrow q^i_\pm }(x_1,x_2,x)$$
$$+f_{q^i_-G_-}(x_1,x_2, Q^2)R_{q^i_-G_-\rightarrow q^i_\pm }(x_1,x_2,x). \eqno(1i)$$

    The explanations about this equation are summarized as follows. (i) We
neglect the linear terms of the AP equation in Eq. 1. (ii) The
first integral in Eqs. 1a and 1e is from two-parton-to-two-parton
amplitudes and it leads to the positive (antishadowing) effect,
while the second integral is the contributions of the interference
amplitudes between the one-parton-to-two-parton and the
three-parton-to-two-parton processes, and it is the negative
(shadowing) effect. The contributions from the virtual diagrams
are cancelled each other in the nucleon. (iii)
$R_{A_mB_n\rightarrow C_l}(x_1,x_2,x)$ are the recombination
functions defined as Fig.1, where $m$ and $n$ are the helicities
of parton $A$ and $B$, and we shall detail them in next Section.
(iii) The nonlinear coefficient $K$ depends on the definition of
the two parton correlation $f_{A_mB_n}(x_1,x_2,Q^2)$ and the
geometric distributions of partons inside the target. For
simplicity, we take $K$ as a free parameter.

    The parton density is a concept defined at twist-2. The parton
correlation function is the generalization of the parton density
beyond the leading twist and it has not yet been determined either
theoretically or experimentally. Similar to [4], we use a toy
model (model I) to assume that

$$(x_1+x_2)f_{A_mB_n}(x_1,x_2,Q^2)=\chi x_1x_2A_m(x_1)B_n(x_2)\theta(1-x_1-x_2),
\eqno(2)$$ where the correlator $\chi$ is regarded as a constant
and it can be incorporated into the free parameter $K$.

    The equation (1) can immediately be simplified using parity
conservation in QCD, which leads to

$$R_{A_+B_+\rightarrow C_+}=R_{A_-B_-\rightarrow C_-}, R_{A_+B_-\rightarrow C_+}
=R_{A_-B_+\rightarrow C_-},$$
$$R_{A_+B_-\rightarrow C_-}=R_{A_-B_+\rightarrow C_+}, R_{A_+B_+\rightarrow C_-}
=R_{A_-B_-\rightarrow C_+}. \eqno(3)$$

    We define

$$A=A_++A_-,~\Delta A=A_+-A_-, \eqno(4)$$
and $A$ is the unpolarized parton (quark or gluon) density. In the
meantime, the following definitions are convenient in the
summation of Eq. (1).

$$R_{AB\rightarrow C}=\frac{1}{2}[R_{A_+B_+\rightarrow C_+}+R_{A_+B_+\rightarrow C_-}
+R_{A_+B_-\rightarrow C_+}+R_{A_+B_-\rightarrow C_-}],\eqno(5a)$$

$$\Delta R^I_{AB\rightarrow C}=\frac{1}{2}[R_{A_+B_+\rightarrow C_+}-R_{A_+B_+\rightarrow
C_-}-R_{A_+B_-\rightarrow C_+}+R_{A_+B_-\rightarrow
C_-}],\eqno(5b)$$

$$\Delta R^{II}_{AB\rightarrow C}=\frac{1}{2}[R_{A_+B_+\rightarrow C_+}-R_{A_+B_+\rightarrow
C_-}+R_{A_+B_-\rightarrow C_+}-R_{A_+B_-\rightarrow
C_-}].\eqno(5c)$$

    The result equation is

$$\frac{dxG(x,Q^2)}{d\ln Q^2}$$
$$=\frac{\alpha_s^2K}{Q^2}
\int_{x_1+x_2\ge x}dx_1dx_2\frac{xx_1x_2}{x_1+x_2}[Eq.6b]$$
$$-2\int_{(x_1+x_2)/2\ge x}dx_1dx_2\frac{xx_1x_2}{x_1+x_2}[Eq.6b],\eqno(6a)$$

$$Eq.6b=G(x_1)G(x_2)\theta(1-x_1-x_2)R_{GG\rightarrow G }(x_1,x_2,x)$$
$$+\sum_iq^i(x_1)\overline{q}^i(x_2)\theta(1-x_1-x_2) R_{q^i\overline{q}^i
\rightarrow G }(x_1,x_2,x)$$
$$+\sum_iG(x_1)q^i(x_2)\theta(1-x_1-x_2)R_{Gq^i\rightarrow G }(x_1,x_2,x),\eqno(6b)$$

$$\frac{dx\Delta G(x,Q^2)}{d\ln Q^2}$$
$$=\frac{\alpha_s^2K}{Q^2}
\int_{x_1+x_2\ge x}dx_1dx_2\frac{xx_1x_2}{x_1+x_2}[Eq.7b]$$
$$-2\int_{(x_1+x_2)/2\ge x}dx_1dx_2\frac{xx_1x_2}{x_1+x_2}[Eq.7b],\eqno(7a)$$

$$Eq.7b=
G(x_1)\Delta G(x_2)\theta(1-x_1-x_2)\Delta R^I_{GG\rightarrow G
}(x_1,x_2,x)$$
$$+\Delta G(x_1)G(x_2)\theta(1-x_1-x_2)\Delta R^{II}_{GG\rightarrow G
}(x_1,x_2,x)$$
$$+\sum_iq^i(x_1)\Delta \overline {q}^i(x_2)\theta(1-x_1-x_2)
\Delta R^I_{q^i\overline{q}^i\rightarrow G }(x_1,x_2,x)$$
$$+\sum_i\Delta q^i(x_1)\overline {q}^i(x_2)\theta(1-x_1-x_2)
\Delta R^{II}_{q^i\overline{q}^i\rightarrow G }(x_1,x_2,x)$$
$$+\sum_iG(x_1)\Delta {q}^i(x_2)\theta(1-x_1-x_2)
\Delta R^I_{Gq^i\rightarrow G }(x_1,x_2,x)$$
$$+\sum_i\Delta G(x_1)q^i(x_2)\theta(1-x_1-x_2)
\Delta R^{II}_{Gq^i\rightarrow G }(x_1,x_2,x), \eqno(7b)$$

$$\frac{dxq^i(x,Q^2)}{d\ln Q^2}$$
$$=\frac{\alpha_s^2K}{Q^2}
\int_{x_1+x_2\ge x}dx_1dx_2\frac{xx_1x_2}{x_1+x_2}[Eq.8b]$$
$$-2\int_{(x_1+x_2)/2\ge x}dx_1dx_2\frac{xx_1x_2}{x_1+x_2}[Eq.8b],\eqno(8a)$$

$$Eq.8b=G(x_1)G(x_2)\theta(1-x_1-x_2)R_{GG\rightarrow q^i}(x_1,x_2,x)$$
$$+q^j(x_1)q^k(x_2)\theta(1-x_1-x_2) R_{q^j q^k
\rightarrow q^i }(x_1,x_2,x)$$
$$+q^i(x_1)G(x_2)\theta(1-x_1-x_2)R_{q^iG\rightarrow q^i}(x_1,x_2,x),\eqno(8b)$$

$$\frac{dx\Delta q^i(x,Q^2)}{d\ln Q^2}$$
$$=\frac{\alpha_s^2K}{Q^2}
\int_{x_1+x_2\ge x}dx_1dx_2\frac{xx_1x_2}{x_1+x_2}[Eq.9b]$$
$$-2\int_{(x_1+x_2)/2\ge x}dx_1dx_2\frac{xx_1x_2}{x_1+x_2}[Eq.9b],\eqno(9a)$$

$$Eq.9b=
G(x_1)\Delta G(x_2)\theta(1-x_1-x_2)\Delta R^I_{GG\rightarrow
q^i}(x_1,x_2,x)$$
$$+\Delta G(x_1)G(x_2)\theta(1-x_1-x_2)\Delta R^{II}_{GG\rightarrow
q^i}(x_1,x_2,x)$$
$$+q^j(x_1)\Delta \overline{q}^j(x_2)\theta(1-x_1-x_2)
\Delta R^I_{q^j\overline{q}^j\rightarrow q^i}(x_1,x_2,x)$$
$$+\Delta q^j(x_1)\overline{q}^j(x_2)\theta(1-x_1-x_2)
\Delta R^{II}_{q^j\overline{q}^j\rightarrow q^i}(x_1,x_2,x)$$
$$+q^j(x_1)\Delta q^k(x_2)\theta(1-x_1-x_2)
\Delta R^I_{q^jq^k\rightarrow q^i}(x_1,x_2,x)$$
$$+\Delta q^j(x_1)q^k(x_2)\theta(1-x_1-x_2)
\Delta R^{II}_{q^jq^k\rightarrow q^i}(x_1,x_2,x)$$
$$+q^i(x_1)\Delta G(x_2)\theta(1-x_1-x_2)
\Delta R^I_{q^iG\rightarrow q^i}(x_1,x_2,x)$$
$$+\Delta q^i(x_1)G(x_2)\theta(1-x_1-x_2)
\Delta R^{II}_{q^iG\rightarrow  q^i}(x_1,x_2,x). \eqno(9b)$$

\newpage
\begin{center}
\section{Polarized recombination functions}
\end{center}

    The parton recombination function for $A(p_1)+B(p_2)\rightarrow C(p_3)+D(p_4)$
in Eq. (1) is defined as

$$\alpha_s^2R_{(p_1p_2\rightarrow p_3)}dx_4\frac{d\textbf{l}_\perp^2}{\textbf{l}_\perp^4}
=\frac{1}{16\pi^2}\frac{x_3x_4}{(x_1+x_2)^3}|M_{p_1p_2\rightarrow
p_3p_4}|^2dx_4\frac{d\textbf{l}_\perp^2}{\textbf{l}_\perp^4},
\eqno(10)$$ where $\textbf{l}_\perp^2=p_x^2+p_y^2$. The matrix
$M_{p_1p_2\rightarrow p_3p_4}$ describe the subprocess (for
example, Fig. 1), which are factorized from the deep inelastic
scattering amplitudes due to the application of TOPT in [3]. Note
that the parton indicated by `$\times$' is on-mass-shell,
(although is off-energy-shell) in TOPT, therefore, it has
determinate helicity. Concretely, the momenta of the initial and
final partons are parameterized as

$$p_1=[x_1p,0,0,x_1p],~  p_2=[x_2p,0,0,x_2p],$$
$$p_3=[x_3p+\frac{p_x^2+p_y^2}{2x_3p},p_x,p_y,x_3p],~;
p_4=[x_4p+\frac{p_x^2+p_y^2}{2x_4p},-p_x,-p_y,x_4p],$$
$$l_L=[(x_4-x_2)p+\frac{p_x^2+p_y^2}{2x_4p},-p_x,-p_y,(x_4-x_2)p],~$$
$$l_R=[(x_4-x_2)p+\frac{p_x^2+p_y^2}{2x_4p},-p_x,-p_y,(x_4-x_2)p]. \eqno(11)$$

    We take the physical axial gauge and the light-like vector $n$
fixes the gauge as $n\cdot A=0$, $A$ being the gluon field. Now a
recombination function for $G_{\pm}G_{\pm}\rightarrow q^i_{\pm}$
in the $t$-channel is computed in following form,

$$R^t_{G_{\pm}G_{\pm}\rightarrow q^i_{\pm}}=
{\langle\frac{1}{12}\rangle}_{color}\frac{x_3x_4}{(x_1+x_2)^3}
Tr[\not{p_4}\gamma_\alpha\not
{l_L}\gamma_\mu\not{p_3}\gamma_\nu(\frac{1\pm\gamma_5}{2})
\not{l_R}\gamma_\beta]
\frac{\epsilon^{\alpha}_{\pm}\epsilon^{\ast\beta}_{\pm}
\epsilon^{\mu}_{\pm}\epsilon^{\ast\nu}_{\pm}}{l_L^2l_R^2},
\eqno(12)$$ where the polarization vector of the collinear initial
gluon is

$$\epsilon_{\pm}=[0,1,{\pm}i,0]/\sqrt{2}. \eqno(13)$$;

$$R^t_{G_{\pm}G_{\pm}\rightarrow G_{\pm}}=
{\langle\frac{9}{8}\rangle}_{color}\frac{x_3x_4}{(x_1+x_2)^3}
C_{\mu\kappa\chi}C_{\nu\zeta\phi}C_{\alpha\eta\rho}C_{\beta\lambda\sigma}
\epsilon_{\pm}^{\mu}\epsilon_{\pm}^{\ast\nu}
\epsilon_{\pm}^{\alpha}\epsilon_{\pm}^{\ast\beta}
\varepsilon_{\pm}^{\kappa}\varepsilon_{\pm}^{\ast\zeta}$$
$$\times \frac{d_{\chi\eta}(l_L)d_{\phi\lambda}(l_R)}{l_L^2l_R^2}
[\delta^{rs}-\frac{p_4^rp_4^s}{|\overrightarrow{p_4}|^2}],\eqno(14)$$
where $r, s$ are the space indices of $\rho, \sigma$ of $p_4$;
$C_{\alpha\eta\rho}$ and $C_{\beta\lambda\sigma}$ are triple gluon
vertex and the polarization vector of final gluon, which has
transverse momentum takes

$$\varepsilon_{\pm}=[0,\frac{1}{\sqrt{2}},\frac{{\pm}i}{\sqrt{2}},
-\frac{p_x\pm ip_y}{\sqrt{2}x_3p}], \eqno(15)$$

The gluon polarization tensor on the Feynman propagator is summing
over helicity state since it is off-shell, i.e.,

$$d_{\mu\nu}(l)=g_{\mu\nu}-\frac{l_\mu n_\nu+l_\nu
n_\mu}{l\cdot n}. \eqno(16)$$

    Similarly, we write other recombination functions in the $t$-chanel

$$R^t_{q^j_{\pm}q^k_{\pm}\rightarrow q^i_{\pm}}=
{\langle\frac{2}{9}\rangle}_{color}\frac{x_3x_4}{(x_1+x_2)^3}
Tr[\not{p_3}\gamma_\mu(\frac{1\pm\gamma_5}{2})\not{
p_1}\gamma_\nu(\frac{1\pm\gamma_5}{2})]$$
$$\times Tr[\not{p_4}\gamma_\alpha\not{p_2}
\gamma_\beta(\frac{1\pm\gamma_5}{2})]
\frac{d_{\mu\alpha}(l_L)d_{\nu\beta}(l_R)}{l_L^2l_R^2}.
\eqno(17)$$

$$R^t_{q^i_{\pm}G_{\pm}\rightarrow q^i_{\pm}}=
{\langle\frac{1}{2}\rangle}_{color}\frac{x_3x_4}{(x_1+x_2)^3}
Tr[\not{p_3}\gamma_\mu(\frac{1\pm\gamma_5}{2})\not{
p_1}\gamma_\nu(\frac{1\pm\gamma_5}{2})]$$
$$\times C_{\alpha\eta\rho}C_{\beta\lambda\sigma}
\epsilon_{\pm}^{\alpha}\epsilon_{\pm}^{\ast\beta}
\frac{d_{\mu\eta}(l_L)d_{\nu\lambda}(l_R)}{l_L^2l_R^2}
[\delta^{rs}-\frac{p_4^rp_4^s}{|\overrightarrow{p_4}|^2}],
\eqno(18)$$

$$R^t_{G_{\pm}q^i_{\pm}\rightarrow G_{\pm}}=
{\langle\frac{1}{2}\rangle}_{color}\frac{x_3x_4}{(x_1+x_2)^3}
Tr[\not{p_4}\gamma_\mu\not{
p_2}\gamma_\nu(\frac{1\pm\gamma_5}{2})]$$
$$\times C_{\alpha\eta\rho}C_{\beta\lambda\sigma}
\epsilon_{\pm}^{\alpha}\epsilon_{\pm}^{\ast\beta}
\varepsilon_{\pm}^{\rho}\varepsilon_{\pm}^{\ast\sigma}
\frac{d_{\mu\eta}(l_L)d_{\nu\lambda}(l_R)}{l_L^2l_R^2},
\eqno(19)$$ where the polarization vector for the on-shell final
state gluon ($p_4$), which has transverse components is

$$\varepsilon_{\pm}=[0,\frac{1}{\sqrt{2}},\frac{{\pm}i}{\sqrt{2}},
\frac{p_x\pm ip_y}{\sqrt{2}x_4p}]. \eqno(20)$$

$$R^t_{q^j_{\pm}\overline{q}^j_{\pm}\rightarrow G_{\pm}}=
{\langle\frac{16}{27}\rangle}_{color}\frac{x_3x_4}{(x_1+x_2)^3}
Tr[\not{p_2}\gamma_\alpha(\frac{1\pm\gamma_5}{2})\not{
l_L}\gamma_\mu\not{
p_1}(\frac{1\pm\gamma_5}{2})\gamma_\nu\not{l_R}\gamma_\beta ]$$
$$\times \frac{\varepsilon_{\pm}^{\mu}\varepsilon_{\pm}^{\ast\nu}}
{l_L^2l_R^2}[\delta^{rs}-\frac{p_4^rp_4^s}{|\overrightarrow{p_4}|^2}].
\eqno(21)$$

    The massless partons with the parallel momenta can go
on-mass-shell simultaneously in the collinear case and the
collinear singularity may arise in the $s$-channel in calculations
of the recombination function. In this case, the collinear time
ordered perturbation theory, which is developed in [10] provides a
safe way to evolute such channel.  For example,

$$R^s_{G_{\pm}G_{\pm}\rightarrow q^i_{\pm}}=
{\langle\frac{3}{16}\rangle}_{color}\frac{x_3x_4}{(x_1+x_2)^3}
Tr[\not{p_3}\gamma_\rho(\frac{1\pm\gamma_5}{2})
\not{p_4}\gamma_\sigma]\frac{n^\eta n^\rho}{(l_L\cdot
n)^2}\frac{n^\lambda n^\sigma}{(l_R\cdot n)^2}
$$
$$\times C^{\mu\alpha\eta}C^{\nu\beta\lambda}\epsilon_\pm^\mu
\epsilon_\pm^{\ast\nu}\epsilon_\pm^\alpha
\epsilon_\pm^{\ast\beta},\eqno(23)$$

$$R^s_{G_{\pm}G_{\pm}\rightarrow G_{\pm}}=
{\langle\frac{9}{8}\rangle}_{color}\frac{x_3x_4}{(x_1+x_2)^3}
\frac{n^\eta n^\rho}{(l_L\cdot
n)^2}C^{\mu\alpha\eta}C^{\kappa\chi\rho}\frac{n^\lambda
n^\sigma}{(l_R\cdot n)^2}C^{\zeta\phi\sigma}C^{\nu\beta\lambda}
$$
$$\times\epsilon_\pm^\mu
\epsilon_\pm^{\ast\nu}\epsilon_\pm^\alpha
\epsilon_\pm^{\ast\beta}\varepsilon_{\pm}^{\rho}
\varepsilon_{\pm}^{\ast\sigma} [\delta^{rs}-\frac{p_4^rp_4^s}
{|\overrightarrow{p_4}|^2}]\eqno(24)$$

$$R^s_{q^j_{\pm}\bar{q}^j_{\pm}\rightarrow q^i_{\pm}}=
{\langle\frac{2}{9}\rangle}_{color}\frac{x_3x_4}{(x_1+x_2)^3}
Tr[\not{p_4}\gamma_\alpha\not{p_3}\gamma_\beta(\frac{1\pm\gamma_5}{2})]
Tr[\not{p_2}\gamma_\mu(\frac{1\pm\gamma_5}{2})
\not{p_1}\gamma_\nu(\frac{1\pm\gamma_5}{2})]$$$$\times\frac{n^\mu
n^\alpha}{(l_L\cdot n)^2}\frac{n^\nu n^\beta}{(l_R\cdot n)^2},
\eqno(25)$$

$$R^s_{q^i_{\pm}G_{\pm}\rightarrow q^i_{\pm}}=
{\langle\frac{2}{9}\rangle}_{color}\frac{x_3x_4}{(x_1+x_2)^3}
Tr[\not{p_3}\gamma_\alpha(\frac{1\pm\gamma_5}{2})\frac{\gamma\cdot
n}{2l_L\cdot
n}\gamma_\mu\not{p_1}\gamma_\nu(\frac{1\pm\gamma_5}{2})\frac{\gamma\cdot
n}{2l_R\cdot n}\gamma_\beta]$$
$$\times\epsilon_{\pm}^{\mu}\epsilon_{\pm}^{\ast\nu}
[\delta^{rs}-\frac{p_4^rp_4^s}{|\overrightarrow{p_4}|^2}],
\eqno(26)$$

$$R^s_{G_{\pm}q^i_{\pm}\rightarrow G_{\pm}}=
{\langle\frac{2}{9}\rangle}_{color}\frac{x_3x_4}{(x_1+x_2)^3}
Tr[\not{p_4}\gamma_\alpha\frac{\gamma\cdot n}{2l_L\cdot
n}\gamma_\mu\not{p_2}\gamma_\nu(\frac{1\pm\gamma_5}{2})\frac{\gamma\cdot
n}{2l_R\cdot n}\gamma_\beta]$$
$$\times\epsilon_{\pm}^{\mu}\epsilon_{\pm}^{\ast\nu}
\varepsilon_{\pm}^{\alpha}\varepsilon_{\pm}^{\ast\beta},
\eqno(27)$$ and

$$R^s_{q^j_{\pm}\bar{q}^j_{\pm}\rightarrow G_{\pm}}=
{\langle\frac{4}{3}\rangle}_{color}\frac{x_3x_4}{(x_1+x_2)^3}
Tr[\not{p_2}\gamma_\rho(\frac{1\pm\gamma_5}{2})
\not{p_1}\gamma_\sigma(\frac{1\pm\gamma_5}{2})]\frac{n^\eta
n^\rho}{(l_L\cdot n)^2}\frac{n^\lambda n^\sigma}{(l_R\cdot n)^2}
$$
$$\times C^{\mu\alpha\eta}C^{\nu\beta\lambda}
\varepsilon_{\pm}^{\mu}\varepsilon_{\pm}^{\ast\nu}
[\delta^{rs}-\frac{p_4^rp_4^s}{|\overrightarrow{p_4}|^2}].
\eqno(28)$$

 The complete polarized recombination functions, which
summing the contributions of the $t$-, $u$-, $s$- and their
interference channels present in Appendix. At the small x limit,
one can keep only the $\ln Q^2 \ln (1/x)$ factor (i.e., the $DLLA$
) and the gluonic initial partons, the non-vanished polarized
recombination functions can be simplified as

$$R_{G_+G_+\rightarrow G_+}=
\frac{9}{4}\frac{1}{x}(\frac{x_1^2}{x_2^2}+\frac{x_2^2}{x_1^2}+1),\eqno(29)$$

$$R_{G_+G_+\rightarrow G_-}= \frac{9}{4}{\frac
{{x_{{1}}}^{4}+{x_{{2}}}^{2}{x_{{1}}}^{2}+{x_{{2}}}^{4}}{x
{x_{{2}}}^{2}{x_{{1}}}^{2}}},\eqno(30)$$ and

$$R_{G_+G_-\rightarrow G_+}= R_{G_+G_-\rightarrow G_-}=
\frac{9}{4}{\frac
{5\,{x_{{1}}}^{6}+6\,x_{{2}}{x_{{1}}}^{5}+3\,{x_{{2}}}^{2}
{x_{{1}}}^{4}-4\,{x_{{2}}}^{3}{x_{{1}}}^{3}+3\,{x_{{2}}}^{4}{x_{{1}}}^
{2}+6\,{x_{{2}}}^{5}x_{{1}}+5\,{x_{{2}}}^{6}}{x \left(
x_{{1}}+x_{{2}} \right)
^{2}{x_{{2}}}^{2}{x_{{1}}}^{2}}}.\eqno(31)$$

    Further assuming $x_1=x_2$, we have

$$R_{G_+G_+\rightarrow G_+}=
R_{G_+G_+\rightarrow G_-}= \frac{27}{4}\frac{1}{x}, \eqno(32)$$

$$R_{G_+G_-\rightarrow G_+}= R_{G_+G_-\rightarrow G_-}={\frac
{27}{2}}\frac{1}{x}. \eqno(33)$$ In this case, Eqs. (5b), (5c) and
(7) are zero. It implies that the polarized recombination effect
$disappears$.

    The corresponding unpolarized gluon recombination functions
at the $DLLA$ are

$$R_{GG\rightarrow G}=\frac{9}{4}{\frac
{6\,{x_{{1}}}^{6}+8\,x_{{2}}{x_{{1}}}^{5}+5\,{x_{{2}}}^{2}
{x_{{1}}}^{4}-2\,{x_{{2}}}^{3}{x_{{1}}}^{3}+5\,{x_{{2}}}^{4}{x_{{1}}}^
{2}+8\,{x_{{2}}}^{5}x_{{1}}+6\,{x_{{2}}}^{6}}{ \left(
x_{{1}}+x_{{2}}
 \right) ^{2}{x_{{2}}}^{2}{x_{{1}}}^{2}x}}\eqno(34)$$
for $x_1\ne x_2$, and

$$R_{GG\rightarrow G}=\frac{81}{4}\frac{1}{x},\eqno(35)$$
for $x_1=x_2$.

\newpage
\begin{center}
\section{Discussions and summary}
\end{center}

    Comparing with the linear spin-dependent AP equation, the
nonlinear recombination equation has more complicated structure,
in particulary this equation contains yet unknown twist-4 matrix,
which is defined as the correlation function
$f_{A_mB_n}(x_1,x_2,Q^2)$ in Eq. (1). The concrete form about the
correlation function is an unresolved problem. Although we used a
toy model (Eq. (2)), we still have other choices. For example,

    Model II.
$$f_{A_{\pm},B_{\mp}}(x_1,x_2,Q^2)\gg
f_{A_{\pm},B_{\pm}}(x_1,x_2,Q^2), \eqno(36)$$ i.e., the
recombination is dominated by two polarized partons with opposite
helicity.  In this case, we have

$$\Delta R^I_{AB\rightarrow CD}=-\Delta R^{II}_{AB\rightarrow CD},
\eqno(37)$$ and

$$R_{AB\rightarrow CD}\rightarrow \frac{1}{2}[R_{A_+B_-\rightarrow
C_+D}+R_{A_-B_+\rightarrow C_-D}]. \eqno(38)$$

    Model III.
$$f_{A_{\pm},B_{\pm}}(x_1,x_2,Q^2)\gg
f_{A_{\pm},B_{\mp}}(x_1,x_2,Q^2), \eqno(39)$$ i.e., the
recombination partons have same helicity. In this case, we have

$$\Delta R^I_{AB\rightarrow CD}=\Delta R^{II}_{AB\rightarrow CD},
\eqno(40)$$ and

$$R_{AB\rightarrow CD}\rightarrow \frac{1}{2}[R_{A_+B_+\rightarrow
C_+D}+R_{A_+B_+\rightarrow C_-D}]. \eqno(41)$$

    One can find the singularities in the recombination equation
(6)-(7), which arise from soft initial partons.

 (Such singularities also exist in the derivation of the GLR-MQ
equation in [8], where an unusual $i\epsilon$ prescription and the
contour deformation of integral are used to avoid the pole.
However, we can not use similar technique to exclude IR
singularity in Eq. (6)-(7) in the case $x_1\ne x_2$. The reason is
that if $x_1$ or $x_2$ in the parton recombination functions is
integrated by contour integral, the definition
$f_{A_mB_n}(x_1,x_2,Q^2)\sim A_m(x_1)B_n(x_2)$ becomes
unreasonable.)

    Since soft partons between the perturbative and
nonperturbative parts of a DIS amplitude should be absorbed into
the nonperturbative correlation function in the collinear
factorization scheme, we further modify the above mentioned
correlation models to avoid the contributions from soft initial
partons. A such correlator between two initial partons with
different values of $x$ is used in [11]. In concrete, we assume
that a parton with rapidity $y=\ln(1/x_1)$ combines with an other
parton with rapidity $y+\eta=\ln(1/x_2)$, where $x_2=\xi x_1$ and
$\xi=e^{-\eta}$, one can generalized the correlation function Eq.
(2) to

$$(x_1+x_2)f_{A_mB_n}(x_1,x_2,Q^2)=\chi(\vert \eta\vert)
x_1x_2A_m(x_1)B_n(x_2)\theta(1-x_1-x_2),\eqno(42)$$ where
$\chi(\vert \eta\vert)$ is a normalized correlator of fusing
partons with the rapidity difference $\eta$. For example, we can
take

$$\chi(\vert \eta\vert)=\frac{a}{\sqrt{\pi}}e^{-a\eta^2}.
\eqno(43)$$ If $a=5$, it implies that $\vert
x_2-x_1\vert_{max}=0.0067$.

     In summary, the corrections of parton recombination to the
polarized AP evolution equation are derived. The relation between
spin-dependent and spin-independent recombination functions and
corresponding evolution equations are discussed.

\vspace{0.3cm}

\noindent {\bf Acknowledgments}: We would like to thank Y.D. Li
and J.F. Yang for discussions.  This work was supported by
following National Natural Science Foundations of China 10075020,
90103013, 10135060 and 10205004.

\newpage
\noindent {\bf Appendix}:

    For completeness we list here the results for the polarized
recombination functions as calculated using equations in Section.
3.

\begin{eqnarray*}
&& G_+G_+\rightarrow G_+ =\frac{9}{4}{\frac { \left(
x_{{1}}+x_{{2}}-x \right) ^{3}}{x{x_{{2}}}^{ 2} \left(
x_{{1}}+x_{{2}} \right) ^{3}{x_{{1}}}^{2}} } ({x_{{1}}}^{
4}-2\,{x_{{1}}}^{3}x+{x_{{1}}}^{2}{x}^{2}+{x_{{2}}}^{4}-2\,{x_{{2}}}^{
3}x+{x_{{2}}}^{2}{x}^{2}\nonumber \\
&&\hspace{30mm}+{x_{{1}}}^{2}{x_{{2}}}^{2}-{x_{{1}}}^{2}x_{{2
}}x-x_{{1}}{x_{{2}}}^{2}x+x_{{1}}x_{{2}}{x}^{2} )\hspace{50mm}\rlap{(A.1)}\\
&& G_+G_+\rightarrow G_- =\frac{9}{4}\frac { \left(
x_{{1}}+x_{{2}}-x \right)}{x{x_{{2}}}^{2} \left( x_{{1}}+x _{{2}}
\right) ^{3}{x_{{1}}}^{2}} (6\,{x_{{1}}}^{4
}x_{{2}}x+6\,{x_{{1}}}^{3}{x_{{2}}}^{2}x-3\,{x_{{1}}}^{3}x_{{2}}{x}^{2
}-7\,{x_{{1}}}^{2}x_{{2}}{x}^{3}\nonumber \\
&&\hspace{30mm}+11\,{x_{{1}}}^{2}{x_{{2}}}^{2}{x}^{2}
+6\,{x_{{2}}}^{4}x_{{1}}x+6\,{x_{{2}}}^{3}{x_{{1}}}^{2}x-3\,{x_{{2}}}^
{3}x_{{1}}{x}^{2}-7\,{x_{{2}}}^{2}x_{{1}}{x}^{3}+2\,x_{{1}}x_{{2}}{x}^
{4}\nonumber \\
&&\hspace{30mm}+{x_{{1}}}^{6}+{x_{{2}}}^{6}+2\,{x_{{1}}}^{5}x_{{2}}+2\,{x_{{1}}}^{
4}{x_{{2}}}^{2}-{x_{{1}}}^{4}{x}^{2}-2\,{x_{{1}}}^{3}{x}^{3}+2\,{x_{{1
}}}^{2}{x}^{4}+2\,{x_{{2}}}^{5}x_{{1}}\nonumber \\
&&\hspace{30mm}+2\,{x_{{2}}}^{4}{x_{{1}}}^{2}-{
x_{{2}}}^{4}{x}^{2}-2\,{x_{{2}}}^{3}{x}^{3}+2\,{x_{{2}}}^{2}{x}^{4}+2
\,{x_{{1}}}^{3}{x_{{2}}}^{3})\hspace{30mm}\rlap{(A.2)}
\\
&& G_+G_-\rightarrow G_+ ={\frac {9}{4}}\,{\frac
{(x_{{1}}+x_{{2}}-x)}{x{x_{{2}}}^ {2}\left (x_{{1}}+x_{{2}}\right
)^{7}{x_{{1}}}^{2}}}(141\,
{x_{{1}}}^{7}{x}^{2}x_{{2}}+42\,{x_{{1}}}^{4}{x}^{4}{x_{{2}}}^{2}-100
\,{x_{{1}}}^{8}xx_{{2}}\nonumber\\
 &&\hspace{30mm}+19\,{x_{{1}}}^{5}{x}^{4}x_{{2}}-5\,{x_{{2}}}^{
8}x_{{1}}x+39\,{x_{{1}}}^{2}{x}^{4}{x_{{2}}}^{4}+137\,{x_{{1}}}^{3}{x_
{{2}}}^{6}x-86\,{x_{{1}}}^{6}{x}^{3}x_{{2}}\nonumber\\
&&\hspace{30mm} +26\,{x_{{2}}}^{9}x_{{1}}+5
\,{x_{{1}}}^{10}+5\,{x_{{2}}}^{10}+155\,{x_{{1}}}^{4}{x}^{2}{x_{{2}}}^
{4}-40\,{x_{{1}}}^{2}{x_{{2}}}^{5}{x}^{3}-212\,{x_{{1}}}^{6}x{x_{{2}}}
^{3}\nonumber\\
&&\hspace{30mm}
-124\,{x_{{1}}}^{3}{x}^{3}{x_{{2}}}^{4}-196\,{x_{{1}}}^{4}{x}^{3}{
x_{{2}}}^{3}-79\,{x_{{1}}}^{2}{x_{{2}}}^{6}{x}^{2}+128\,{x_{{1}}}^{4}{
x_{{2}}}^{5}x+44\,{x_{{2}}}^{7}{x_{{1}}}^{2}x\nonumber\\
&&\hspace{30mm}-47\,{x_{{1}}}^{5}x{x_{{2
}}}^{4}-177\,{x_{{1}}}^{5}{x}^{3}{x_{{2}}}^{2}-209\,{x_{{1}}}^{7}x{x_{
{2}}}^{2}+40\,{x_{{1}}}^{3}{x}^{4}{x_{{2}}}^{3}-33\,{x_{{1}}}^{3}{x_{{
2}}}^{5}{x}^{2}\nonumber\\
&&\hspace{30mm}+{x_{{2}}}^{6}x_{{1}}{x}^{3}+319\,{x_{{1}}}^{5}{x}^{2}{
x_{{2}}}^{3}+291\,{x_{{1}}}^{6}{x}^{2}{x_{{2}}}^{2}-35\,{x_{{2}}}^{7}x
_{{1}}{x}^{2}+13\,{x_{{2}}}^{5}{x}^{4}x_{{1}}\nonumber\\
&&\hspace{30mm}+64\,{x_{{2}}}^{7}{x_{{1}
}}^{3}+57\,{x_{{2}}}^{8}{x_{{1}}}^{2}+26\,{x_{{1}}}^{9}x_{{2}}+57\,{x_
{{1}}}^{8}{x_{{2}}}^{2}+64\,{x_{{1}}}^{7}{x_{{2}}}^{3}+34\,{x_{{1}}}^{
6}{x_{{2}}}^{4}\nonumber\\
&&\hspace{30mm}+12\,{x_{{1}}}^{5}{x_{{2}}}^{5}+34\,{x_{{1}}}^{4}{x_{{2
}}}^{6}-4\,{x_{{2}}}^{9}x+2\,{x_{{2}}}^{6}{x}^{4}+2\,{x_{{2}}}^{7}{x}^
{3}-5\,{x_{{2}}}^{8}{x}^{2}+30\,{x_{{1}}}^{8}{x}^{2}\nonumber\\
&&\hspace{30mm}-20\,{x_{{1}}}^{7}
{x}^{3}+5\,{x_{{1}}}^{6}{x}^{4}-20\,{x_{{1}}}^{9}x )
\hspace{60mm}\rlap{(A.3)}
\\
&&G_+G_-\rightarrow G_-={\frac {9}{4}}\,{\frac {
(x_{{1}}+x_{{2}}-x )}{x{x_{{2}} }^{2} (x_{{1}}+x_{{2}}
)^{7}{x_{{1}}}^{2}}}(-31\,
{x_{{1}}}^{7}{x}^{2}x_{{2}}+27\,{x_{{1}}}^{4}{x}^{4}{x_{{2}}}^{2}-7\,{
x_{{1}}}^{8}xx_{{2}}+9\,{x_{{1}}}^{5}{x}^{4}x_{{2}}\nonumber\\
&&\hspace{30mm}-104\,{x_{{2}}}^{8}
x_{{1}}x+54\,{x_{{1}}}^{2}{x}^{4}{x_{{2}}}^{4}-206\,{x_{{1}}}^{3}{x_{{
2}}}^{6}x+3\,{x_{{1}}}^{6}{x}^{3}x_{{2}}+26\,{x_{{2}}}^{9}x_{{1}}+5\,{
x_{{1}}}^{10}\nonumber\\
&&\hspace{30mm}+5\,{x_{{2}}}^{10}+115\,{x_{{1}}}^{4}{x}^{2}{x_{{2}}}^{4}
-211\,{x_{{1}}}^{2}{x_{{2}}}^{5}{x}^{3}+125\,{x_{{1}}}^{6}x{x_{{2}}}^{
3}-192\,{x_{{1}}}^{3}{x}^{3}{x_{{2}}}^{4}\nonumber\\
&&\hspace{30mm}-92\,{x_{{1}}}^{4}{x}^{3}{x_{
{2}}}^{3}+319\,{x_{{1}}}^{2}{x_{{2}}}^{6}{x}^{2}-25\,{x_{{1}}}^{4}{x_{
{2}}}^{5}x-217\,{x_{{2}}}^{7}{x_{{1}}}^{2}x+136\,{x_{{1}}}^{5}x{x_{{2}
}}^{4}\nonumber\\
&&\hspace{30mm}-24\,{x_{{1}}}^{5}{x}^{3}{x_{{2}}}^{2}+34\,{x_{{1}}}^{7}x{x_{{2}
}}^{2}+36\,{x_{{1}}}^{3}{x}^{4}{x_{{2}}}^{3}+307\,{x_{{1}}}^{3}{x_{{2}
}}^{5}{x}^{2}-106\,{x_{{2}}}^{6}x_{{1}}{x}^{3}\nonumber\\
&&\hspace{30mm}-41\,{x_{{1}}}^{5}{x}^{2
}{x_{{2}}}^{3}-67\,{x_{{1}}}^{6}{x}^{2}{x_{{2}}}^{2}+157\,{x_{{2}}}^{7
}x_{{1}}{x}^{2}+27\,{x_{{2}}}^{5}{x}^{4}x_{{1}}+64\,{x_{{2}}}^{7}{x_{{
1}}}^{3}\nonumber\\
&&\hspace{30mm}+57\,{x_{{2}}}^{8}{x_{{1}}}^{2}+26\,{x_{{1}}}^{9}x_{{2}}
+57\,{
x_{{1}}}^{8}{x_{{2}}}^{2}+64\,{x_{{1}}}^{7}{x_{{2}}}^{3}+34\,{x_{{1}}}
^{6}{x_{{2}}}^{4}+12\,{x_{{1}}}^{5}{x_{{2}}}^{5}\nonumber\\
&&\hspace{30mm}+34\,{x_{{1}}}^{4}{x_{
{2}}}^{6}-20\,{x_{{2}}}^{9}x+5\,{x_{{2}}}^{6}{x}^{4}-20\,{x_{{2}}}^{7}
{x}^{3}+30\,{x_{{2}}}^{8}{x}^{2}-5\,{x_{{1}}}^{8}{x}^{2}+2\,{x_{{1}}}^
{7}{x}^{3}\nonumber\\
&&\hspace{30mm}+2\,{x_{{1}}}^{6}{x}^{4}-4\,{x_{{1}}}^{9}x)\hspace{82mm}\rlap{(A.4)}
\\
&&q_+^j\bar{q}_+^j\rightarrow G_+ ={\frac {16}{27}}\,{\frac {
(x_{{1}}+x_{{2}}-x )}{x_{{2} }(x_{{1}}+x_{{2}})^{3}xx_{{1}}}} (8\,
{x_{{1}}}^{2}{x_{{2}}}^{2}+4\,{x_{{1}}}^{4}+4\,{x_{{2}}}^{4}+{x_{{2}}}
^{3}x_{{1}}+x_{{2}}{x_{{1}}}^{3}-x_{{2}}{x_{{1}}}^{2}x )
\\
&&\hspace{146mm}\rlap{(A.5)}
 \\
&&q_+^j\bar{q}_+^j\rightarrow G_- ={\frac {16}{27}}\,{\frac {\left
(x_{{1}}+x_{{2}}-x\right )}{x_{{2}}\left (x_{{1}}+x_{{2}}\right
)^{3}xx_{{1}}}}(4\,
{x_{{1}}}^{4}-8\,{x_{{1}}}^{3}x+8\,{x_{{1}}}^{2}{x_{{2}}}^{2}+4\,{x_{{
1}}}^{2}{x}^{2}+4\,{x_{{2}}}^{4}-8\,{x_{{2}}}^{3}x\nonumber\\
&&\hspace{30mm}+4\,{x_{{2}}}^{2}{x}
^{2}+{x_{{2}}}^{3}x_{{1}}-{x_{{2}}}^{2}x_{{1}}x+x_{{2}}{x_{{1}}}^{3}
 )\hspace{52mm}\rlap{(A.6)}
\\
&&q_+^j\bar{q}_-^j\rightarrow G_+ =0\hspace{118mm}\rlap{(A.7)}
\\
&&q_+^j\bar{q}_-^j\rightarrow G_- =0\hspace{118mm}\rlap{(A.8)}
\\
&&G_+q_+^i\rightarrow G_+ =\frac{2}{9}{\frac { (x_{{1}}+x_{{2}}-x
)^{2}}{x_{{1}}(x_{{1}}+x_{{2}})^{3}x{x_{{2}}}^{2}}}(9\,{x}^{2}{x_{
{2}}}^{2}-18\,{x_{{2}}}^{3}x+9\,{x_{{2}}}^{4}+4\,{x_{{1}}}^{4})\hspace{25mm}\rlap{(A.9)}
\\
&&G_+q_+^i\rightarrow G_- =\frac{1}{18}{\frac { (x_{{1}}+x_{{2}}-x
)^{2}}{x_{{1}}(x_{{1}}+x_{{2}})^{4}x {x_{{2}}}^{2}}}(9\,{x_{{2}}}^
{2}{x}^{2}x_{{1}}+9\,{x_{{2}}}^{3}{x}^{2}+18\,{x_{{2}}}^{3}xx_{{1}}+36
\,{x_{{2}}}^{4}x+36\,{x_{{2}}}^{4}x_{{1}}\nonumber\\
&&\hspace{30mm}+36\,{x_{{2}}}^{5}+16\,{x_{{1
}}}^{5}+48\,{x_{{1}}}^{4}x_{{2}}+48\,{x_{{1}}}^{3}{x_{{2}}}^{2}-32\,{x
_{{1}}}^{4}x-78\,{x_{{1}}}^{3}x_{{2}}x+16\,{x_{{1}}}^{2}{x_{{2}}}^{3}\nonumber\\
&&\hspace{30mm}-
68\,{x_{{1}}}^{2}{x_{{2}}}^{2}x+16\,{x_{{1}}}^{3}{x}^{2}+12\,{x_{{1}}}
^{2}{x}^{2}x_{{2}})\hspace{50mm}\rlap{(A.10)}
\\
&&G_+q_-^i\rightarrow G_+ =\frac{1}{18}{\frac {\left
(x_{{1}}+x_{{2}}-x\right )^{2}}{{x_{{1}}(x_{{1}}+x_{{2}}
)^{5}{x_{{2}}}^{2}x}}}(7\,{x}^{2}{x_
{{1}}}^{2}{x_{{2}}}^{2}+18\,{x}^{2}{x_{{2}}}^{3}x_{{1}}+9\,{x}^{2}{x_{
{2}}}^{4}+16\,{x_{{1}}}^{6}+64\,{x_{{1}}}^{5}x_{{2}}\nonumber\\
&&\hspace{30mm}+132\,{x_{{1}}}^{4
}{x_{{2}}}^{2}+136\,{x_{{1}}}^{3}{x_{{2}}}^{3}+52\,{x_{{1}}}^{2}{x_{{2
}}}^{4}-18\,{x_{{1}}}^{3}x_{{2}}{x}^{2}-18\,{x_{{1}}}^{2}{x_{{2}}}^{3}
x+18\,{x_{{1}}}^{4}x_{{2}}x )\nonumber\\
&&\hspace{145mm}\rlap{(A.11)}
\\
&&G_+q_-^i\rightarrow G_- =\frac{2}{9}{\frac {\left
(x_{{1}}+x_{{2}}-x\right )^{2}x_{{1}}}{x (x_{{1}}+x_{{2}}
)^{3}{x_{{2}}}^{2}}}(4\,{x_{
{1}}}^{2}-8\,x_{{1}}x+4\,{x}^{2}+9\,{x_{{2}}}^{2}-9\,x_{{2}}x )
 \hspace{18mm}\rlap{(A.12)}
\\
&&G_+G_+\rightarrow q_+^i =\frac{1}{12}\frac { (x_{{1}}+x_{{2}}-x
)^{2}}{(x_{{1}}+x_{{2}} )^{3}{x_{{2}}}^{2}{x_{{1 }}}^{2}}
(4\,{x_{{1}}}
^{4}+7\,{x_{{1}}}^{3}x_{{2}}-8\,{x_{{1}}}^{3}x+2\,{x_{{1}}}^{2}{x_{{2}
}}^{2}-6\,{x_{{1}}}^{2}xx_{{2}}\nonumber\\
&&\hspace{30mm}+4\,{x}^{2}{x_{{1}}}^{2}+4\,{x}^{2}{x_{
{2}}}^{2}-x_{{1}}{x_{{2}}}^{3}+2\,x_{{1}}x{x_{{2}}}^{2}-x_{{1}}x_{{2}}
{x}^{2}) \hspace{28mm}\rlap{(A.13)}
\\
&&G_+G_+\rightarrow q_-^i =\frac{1}{12} \frac { (x_{{1}}+x_{{2}}-x
)^{2}}{(x_{{1}}+x_{{2}} )^{3}{x_{{2}}}^{2}{x_{{1}}}^{2}}
(4\,{x}^{2}{x
_{{1}}}^{2}+4\,{x_{{1}}}^{2}{x_{{2}}}^{2}+8\,x_{{1}}{x_{{2}}}^{3}-8\,x
_{{1}}x{x_{{2}}}^{2}+4\,{x_{{2}}}^{4}\nonumber\\
&&\hspace{30mm} -8\,{x_{{2}}}^{3}x+4\,{x}^{2}{x_{
{2}}}^{2}-x_{{1}}x_{{2}}{x}^{2}) \hspace{65mm}\rlap{(A.14)}
\\
&&G_+G_-\rightarrow q_+^i =\frac{1}{12}\frac { (x_{{1}}+x_{{2}}-x
)^{2}}{ (x _{{1}}+x_{{2}} )^{7}{x_{{2}}}^{2}{x_{{1}}}^{2}}
(4\,{x_{{1}}}
^{6}{x}^{2}+4\,{x_{{1}}}^{4}{x_{{2}}}^{4}-8\,{x_{{1}}}^{7}x+24\,{x_{{1
}}}^{6}{x_{{2}}}^{2}+16\,{x_{{1}}}^{5}{x_{{2}}}^{3}\nonumber\\
&&\hspace{30mm}+16\,{x_{{1}}}^{7}x
_{{2}}+4\,{x_{{1}}}^{8}+24\,{x}^{2}{x_{{2}}}^{5}x_{{1}}+4\,{x}^{2}{x_{
{2}}}^{6}-10\,{x_{{1}}}^{3}{x}^{2}{x_{{2}}}^{3}+33\,{x_{{1}}}^{2}{x}^{
2}{x_{{2}}}^{4}\nonumber\\
&&\hspace{30mm}-8\,{x_{{1}}}^{4}x{x_{{2}}}^{3}+24\,{x_{{1}}}^{5}{x}^{2
}x_{{2}}+33\,{x_{{1}}}^{4}{x}^{2}{x_{{2}}}^{2}+14\,{x_{{1}}}^{3}x{x_{{
2}}}^{4}-40\,{x_{{1}}}^{6}xx_{{2}}\nonumber\\
&&\hspace{30mm}-53\,{x_{{1}}}^{5}x{x_{{2}}}^{2} -8\,
x{x_{{2}}}^{5}{x_{{1}}}^{2}-9\,x_{{1}}x{x_{{2}}}^{6} )
\hspace{50mm}\rlap{(A.15)}
\\
&&G_+G_-\rightarrow q_-^i =\frac{1}{12}\frac { (x_{{1}}+x_{{2}}-x
)^{2}}{(x_{{1}}+x_{{2}} )^{7}{x_ {{2}}}^{2}{x_{{1}}}^{2}}
(24\,{x_{{2}}
}^{6}{x_{{1}}}^{2}+16\,{x_{{2}}}^{7}x_{{1}}-8\,{x_{{2}}}^{7}x+16\,{x_{
{2}}}^{5}{x_{{1}}}^{3}+4\,{x_{{2}}}^{8}\nonumber\\
&&\hspace{30mm}+4\,{x_{{1}}}^{6}{x}^{2}+4\,{x_
{{1}}}^{4}{x_{{2}}}^{4}+6\,{x}^{2}{x_{{2}}}^{5}x_{{1}}+4\,{x}^{2}{x_{{
2}}}^{6}+8\,{x_{{1}}}^{3}{x}^{2}{x_{{2}}}^{3}+15\,{x_{{1}}}^{2}{x}^{2}
{x_{{2}}}^{4}\nonumber\\
&&\hspace{30mm}-13\,{x_{{1}}}^{4}x{x_{{2}}}^{3}+24\,{x_{{1}}}^{5}{x}^{2}
x_{{2}}+51\,{x_{{1}}}^{4}{x}^{2}{x_{{2}}}^{2}-35\,{x_{{1}}}^{3}x{x_{{2
}}}^{4}+{x_{{1}}}^{5}x{x_{{2}}}^{2}\nonumber\\
&&\hspace{30mm}-35\,x{x_{{2}}}^{5}{x_{{1}}}^{2}-22
\,x_{{1}}x{x_{{2}}}^{6} ) \hspace{70mm}\rlap{(A.16)}
\\
&&q_+^j\bar{q}_+^j\rightarrow q_+^i ={\frac {8}{27}}\frac
{x_{{1}}}{(x_{{1}}+x_{{2}})^ {7}x_{{2}}}
(48\,{x_{{1}}}^{5}x_{{2}}+72\,{x
_{{1}}}^{4}{x_{{2}}}^{2}+48\,{x_{{1}}}^{3}{x_{{2}}}^{3}+12\,{x_{{1}}}^
{2}{x_{{2}}}^{4}-12\,{x_{{1}}}^{5}x\nonumber\\
&&\hspace{30mm}+3\,{x}^{2}{x_{{1}}}^{4}+12\,{x_{{1
}}}^{6}-46\,{x_{{1}}}^{4}xx_{{2}}-64\,{x_{{1}}}^{3}x{x_{{2}}}^{2}-36\,
{x_{{1}}}^{2}x{x_{{2}}}^{3}-4\,x_{{1}}x{x_{{2}}}^{4}\nonumber\\
&&\hspace{30mm}+8\,{x}^{2}{x_{{1}
}}^{3}x_{{2}}+18\,{x_{{2}}}^{2}{x}^{2}{x_{{1}}}^{2}+24\,{x_{{2}}}^{3}{
x}^{2}x_{{1}}-20\,{x_{{2}}}^{2}{x}^{3}x_{{1}}+11\,{x_{{2}}}^{4}{x}^{2}\nonumber\\
&&\hspace{30mm}
-22\,{x_{{2}}}^{3}{x}^{3}+12\,{x_{{2}}}^{2}{x}^{4}+2\,x_{{2}}{x}^{3}{x
_{{1}}}^{2}+2\,{x_{{2}}}^{5}x ) \hspace{40mm}\rlap{(A.17)}
\\
&&q_+^j\bar{q}_+^j\rightarrow q_-^i ={\frac {32}{9}}{\frac
{x_{{1}}x_{{2}}\left (-x_{{1}}-x_{{2}}+x \right
)^{2}{x}^{2}}{\left (x_{{1}}+x_{{2}}\right )^{7}}}
 \hspace{73mm}\rlap{(A.18)}
\\
&&q_+^j\bar{q}_-^j\rightarrow q_+^i =0\hspace{120mm}\rlap{(A.19)}
\\
&&q_+^j\bar{q}_-^j\rightarrow q_-^i =0\hspace{120mm}\rlap{(A.20)}\\
&&q_+^jq_+^k\rightarrow q_+^i ={\frac {8}{9}}\,{\frac {\left
(-2\,x_{{1}}+x\right )^{2}x_{{1}}}{ \left (x_{{1}}+x_{{2}}\right
)^{3}x_{{2}}}} \hspace{90mm}\rlap{(A.21)}
\\
&&q_+^jq_+^k\rightarrow q_-^i =0\hspace{120mm}\rlap{(A.22)}
\\
&&q_+^jq_-^k\rightarrow q_+^i =0 \hspace{120mm}\rlap{(A.23)}
\\
&&q_+^jq_-^k\rightarrow q_-^i={\frac {8}{9}}\,{\frac {\left
(-2\,x_{{1}}+x\right )^{2}x_{{2}}}{x_{{1 }}\left
(x_{{1}}+x_{{2}}\right )^{3}}}
 \hspace{90mm}\rlap{(A.24)}
\\
&&q_+^iG_+\rightarrow q_+^i=\frac {1}{18}\frac {
(x_{{1}}+x_{{2}}-x )}{{x_{{2}}}^{2} (x_{{1}}+x_{{2}} )^{3}x_
{{1}}}(45\,{x_{{1}}}^{4}
-90\,{x_{{1}}}^{3}x+54\,{x_{{1}}}^{3}x_{{2}}-54\,{x_{{1}}}^{2}x_{{2}}x
+45\,{x_{{1}}}^{2}{x}^{2}\nonumber\\
&&\hspace{30mm}+81\,{x_{{1}}}^{2}{x_{{2}}}^{2}+16\,{x}^{2}{x
_{{2}}}^{2}) \hspace{75mm}\rlap{(A.25)}
\\
&&q_+^iG_+\rightarrow q_-^i=0\hspace{117mm}\rlap{(A.26)}
\\
&&q_+^iG_-\rightarrow q_+^i=\frac {1}{18}\frac {
(x_{{1}}+x_{{2}}-x )}{{x_{{2}}}^{2} (x_{{1}}+x_{{2}} )^{5}x_{{1}}}
(-22\,{x_{{2}}}^{5
}x_{{1}}-32\,{x_{{2}}}^{5}x+16\,{x_{{2}}}^{4}{x}^{2}+18\,{x_{{2}}}^{3}
{x}^{2}x_{{1}}+4\,{x_{{2}}}^{4}x_{{1}}x\nonumber\\
&&\hspace{30mm}-58\,{x_{{1}}}^{3}{x_{{2}}}^{3}
-85\,{x_{{1}}}^{2}{x_{{2}}}^{4}+9\,{x_{{1}}}^{4}{x}^{2}-18\,{x_{{1}}}^
{5}x+198\,{x_{{1}}}^{3}x{x_{{2}}}^{2}+198\,{x_{{1}}}^{2}{x_{{2}}}^{3}x\nonumber\\
&&\hspace{30mm}-18\,{x_{{1}}}^{3}{x}^{2}x_{{2}}-20\,{x_{{1}}}^{4}{x_{{2}}}^{2}+18\,{x
_{{1}}}^{4}xx_{{2}}-9\,{x}^{2}{x_{{1}}}^{2}{x_{{2}}}^{2}+9\,{x_{{1}}}^
{6}+16\,{x_{{2}}}^{6} )
 \hspace{4mm}\rlap{(A.27)}
\\
&&q_+^iG_-\rightarrow q_-^i=0\hspace{125mm}\rlap{(A.28)}
\end{eqnarray*}

when $x_1=x_2$,
\begin{eqnarray*}
&&G G \rightarrow G ={\frac {9}{64}}\,{\frac {(2\,x_{{1}}-x )
(140\,{x_
{{1}}}^{2}{x}^{2}-116\,x_{{1}}{x}^{3}+29\,{x}^{4}-48\,{x_{{1}}}^{3}x+
72\,{x_{{1}}}^{4})}{{x_{{1}}}^{5}x}} \hspace{20mm} \rlap{(A.29)}
\\
&&q^j \bar{q}^j \rightarrow G  =\frac {4}{27}{\frac {
(2\,x_{{1}}-x) (18\,{x_{{ 1}}}^{2}-9\,x_{{1}}x+4\,{x}^{2}
)}{{x_{{1}}}^{3}x}} \hspace{68mm}\rlap{(A.30)}
\\
&&G q^i\rightarrow G ={\frac {1}{288}}\,{\frac { (2\,x_{{1}}-x
)^{2} (79\,{ x}^{2}-202\,x_{{1}}x+304\,{x_{{1}}}^{2}
)}{{x_{{1}}}^{4}x}} \hspace{56mm} \rlap{(A.31)}
\\
&&GG\rightarrow q^i ={\frac {1}{96}}\,{\frac {\left
(2\,x_{{1}}-x\right )^{2}\left (18\,{x
_{{1}}}^{2}-21\,x_{{1}}x+14\,{x}^{2}\right )}{{x_{{1}}}^{5}}}
 \hspace{61mm}\rlap{(A.32)}
\\
&&q^j\bar{q}^j\rightarrow q^i={\frac {1}{108}}\,{\frac {\left
(2\,x_{{1}}-x\right )^{2}\left (6\,{x
_{{1}}}^{2}+\,x_{{1}}x+3\,{x}^{2}\right )}{{x_{{1}}}^{5}}}
 \hspace{67mm}\rlap{(A.33)}
\\
&&q^jq^k\rightarrow q^i=\frac{2}{9}{\frac { (2\,x_{{1}}-x
)^{2}}{{x_{{1}}}^{3}}}
 \hspace{108mm}\rlap{(A.34)}
 \\
&&q^iG\rightarrow q^i={\frac {1}{288}}\,{\frac { (2\,x_{{1}}-x )
(140\,{x_ {{1}}}^{2}-52\,x_{{1}}x+65\,{x}^{2} )}{{x_{{1}}}^{4}}}
 \hspace{60mm}\rlap{(A.35)}
\end{eqnarray*}
they consist with the results of Ref. [4].

 \newpage

Figure Captions

Fig. 1 The diagrams contributing to the polarized recombination
functions. Here the shaded part implies all possible QCD-channels
and $\times$ means the probing place.

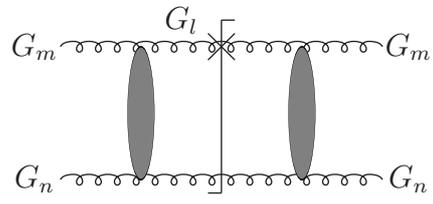
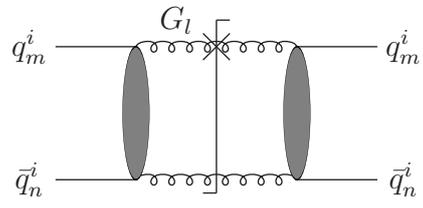
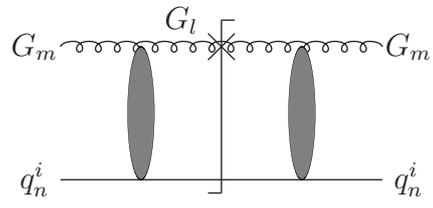
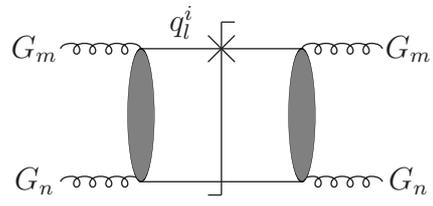
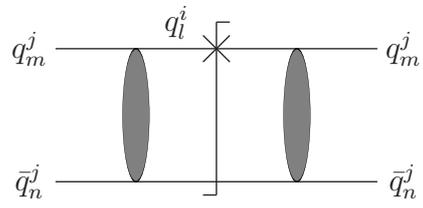
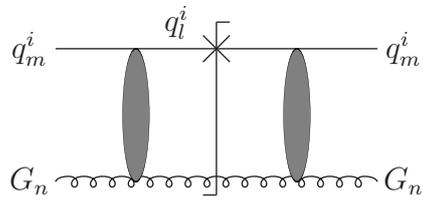
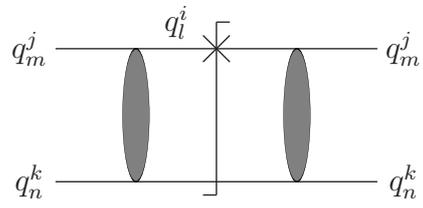
\end{document}